\documentclass[preprint2]{aastex}

\usepackage{graphicx}
\usepackage{amsmath}
\usepackage{natbib}
\usepackage{color}
\usepackage{comment}
\usepackage{rotating}
\usepackage{eso-pic}

\AddToShipoutPictureBG*{%
  \AtPageUpperLeft{%
    \hspace{0.75\paperwidth}%
    \raisebox{-3.5\baselineskip}{%
      \makebox[0pt][l]{\textnormal{DES 2014-0007}}
}}}%

\AddToShipoutPictureBG*{%
  \AtPageUpperLeft{%
    \hspace{0.75\paperwidth}%
    \raisebox{-4.5\baselineskip}{%
      \makebox[0pt][l]{\textnormal{Fermilab PUB-14-578-AE}}
}}}%

\begin{document}
\title{Crowded Cluster Cores: An Algorithm for Deblending in Dark Energy Survey Images}

\author{Yuanyuan Zhang$^1$, Timothy A. McKay$^{1, 2, 3, \star}$ , Emmanuel Bertin$^4$, Tesla Jeltema$^5$, Christopher J. Miller$^{1, 2}$, Eli Rykoff$^6$, Jeeseon Song$^{1,  7}$\vspace{1em}}

\affil{$^1$Physics Department, University of Michigan, Ann Arbor 48109, MI, USA}
\affil{$^2$Astronomy Department, University of Michigan, Ann Arbor 48109, MI, USA}
\affil{$^3$Michigan Center For Theoretical Physics,  University
  of Michigan, Ann Arbor 48109, MI, USA}
\affil{$^4$Institut d'Astrophysique de Paris, Univ. Pierre et Marie Curie \& CNRS
UMR7095, F-75014 Paris, France}
\affil{$^5$Department of Physics and Santa Cruz Institute for Particle Physics, University of California, Santa Cruz, CA, USA}
\affil{$^6$SLAC National Accelerator Laboratory, Menlo Park, CA, USA}
\affil{$^7$Korea Astronomy and Space Science Institute, Daejeon 305-348, Republic of Korea}
\altaffiltext{$\star$}{Following authors listed alphabetically.}

\begin{abstract}
Deep optical images are often crowded with overlapping objects. This is especially true in the cores of galaxy clusters, where images of dozens of galaxies may lie atop one another. Accurate measurements of cluster properties require deblending algorithms designed to automatically extract a list of individual objects and decide what fraction of the light in each pixel comes from each object. In this paper, we introduce new software tool called the Gradient And INterpolation (GAIN) based deblender. GAIN is used as a secondary deblender to improve the separation of overlapping objects in galaxy cluster cores in Dark Energy Survey images. It uses image intensity gradients and an interpolation technique originally developed to correct flawed digital images. This paper is dedicated to describing the algorithm of the GAIN deblender and its applications, but we additionally include modest tests of the software based on real Dark Energy Survey coadd images. GAIN helps to extract unbiased photometry measurement for blended sources and improve detection completeness while introducing few spurious detections. When applied to processed Dark Energy Survey data, GAIN serves as a useful quick fix when a high level of deblending is desired.
\end{abstract}
\maketitle

\section{Introduction}

Deep, wide-field, ground-based imaging surveys have an important role to play in the near future of astronomy. Existing projects like the Pan-STARRS\footnote{\texttt{http://pan-starrs.ifa.hawaii.edu/}} and the Dark Energy Survey (DES)\footnote{\texttt{http://www.darkenergeysurvey.org/}} are mapping thousands of square degrees of the sky to 24th magnitude and beyond. During the next decade, the Large Synoptic Survey Telescope (LSST)\footnote{\texttt{http://www.lsst.org/}} plans to image 20,000 square degrees to 28th magnitude. These surveys support an enormous range of science goals, from identifying near-Earth objects to measuring the expansion history of the universe. One important goal for all of these surveys is to measure the properties of galaxy clusters. Clusters are extreme objects, marking the upper limit of structure formation. As such, properties of their population are highly sensitive to basic cosmological parameters like the expansion history and cosmic mass density. Detecting galaxy clusters and measuring their properties precisely is an important task.

While deep optical images greatly facilitate the detection and measurement of clusters, they present some data processing challenges. There exist several software packages for processing wide-field optical images; each attempts to automatically detect astronomical sources and measure their properties \citep[Lang et al., in prep]{1981AJ.....86..476J, 1990MNRAS.247..311B, 1990MNRAS.243..692M, 1991PASP..103..396Y, 1996A&AS..117..393B, 2000MNRAS.319..700A, 2014arXiv1410.7397L}. The SExtractor \citep{1996A&AS..117..393B} package is among one of the most popular of these tools. These packages have greatly aided astronomical imaging studies, but may become inefficient when processing deep images crowded with objects. A common problem is the failure to detect objects blended with others, i.e., failing to deblend.

The Dark Energy Survey has adopted an advanced version of SExtractor in the data processing pipeline for catalog production (the Dark Energy Survey collaboration, in prep.). To handle crowded images,  SExtractor has a deblending procedure that decides if a detected object should be further separated as several branching components.  Upon application to DES data, a few SExtractor set-ups (including the $DEBLEND\_MINCONT$, $DEBLEND\_NTHRESH$, $CLEAN\_PARAM$ parameters and the image convolution filters) were explored, but the improved detection efficiency of blended sources comes at the cost of an increasing amount of spurious detections and deteriorated photometry measurements. Although an optimum deblending setting has been determined to maximize DES data quality for general purposes, some scientific studies involving crowded regions like galaxy clusters set a different, more demanding requirement. 

The deblending problem also seems to plague other reduction packages. The data processing pipeline of the Sloan Digital Sky Survey (SDSS) is very different from that of DES \citep{2000AJ....120.1579Y, 2001ASPC..238..269L}. However, there are also reports about suppressed completeness around bright objects \citep{2006ApJS..162...38A, 2005MNRAS.361.1287M}, implying that deblending was also an issue. A few other sky survey programs tackle the deblending dilemma with two SExtractor runs, one optimized for detecting isolated objects and one optimized for detecting blended objects, but the output from the second run will need  to be pruned and refined. This is usually achieved with  precision profile fitting methods  (GIM2D Simard et al. 2002; GALFIT Peng
et al. 2002, 2010), which turns out to be too slow for extremely large data sets like those from DES. Another successful practice focuses on deblending images crowded with point sources, like the images of globular clusters \citep{1983Ap&SS..90..405F, 1987PASP...99..191S, 2000A&AS..147..335D, 2007ApJ...661.1339S}, but the technique cannot be directly applied to extra-galactic imaging surveys.  When deblending point sources, the intrinsic shape of every object can be more or less accurately estimated,  but extra-galactic images are dense with galaxies, each with its own unknown shape, brightness, and size. Their images often overlap, making the identification of individual galaxies and measurements of their brightness a challenge.

In this paper, we describe a new software package that aids blended source detection and photometry measurement -- the Gradient And INterpolation based deblender (GAIN deblender).  The GAIN deblender is a secondary package that operates on clean, already processed astronomical images, and requires one round of source extraction to be done before its application. It automatically identifies blended sources, separates them, and prepares the photometry measurement of each individual source. This software is primarily designed for extremely wide surveys like DES, and  running speed is one of the biggest feature. The algorithms may also be of use for other data sets like those from HST and SDSS. We note that a future version of SExtractor featuring an improved deblender is under development (Bertin, private communication), which DES plans to implement upon its delivery. Meanwhile, our software provides a quick fix that assists current DES data production. It can also be combined to use with future SExtractor releases. 

The rest of this paper is organized as follows. We  introduce the features and functions of the GAIN deblender in Section~\ref{sec:app} and explain the algorithms in Section~\ref{sec:method}. We then analyze the effectiveness of this approach  with real full-depth DES data, and the result is presented in Section~\ref{sec:veri}. The software characteristics are summarized and discussed in Section \ref{sec:sum}.

\section{Software Features and Functions}

\label{sec:app}
\begin{figure*}
\includegraphics[width=1.0\textwidth]{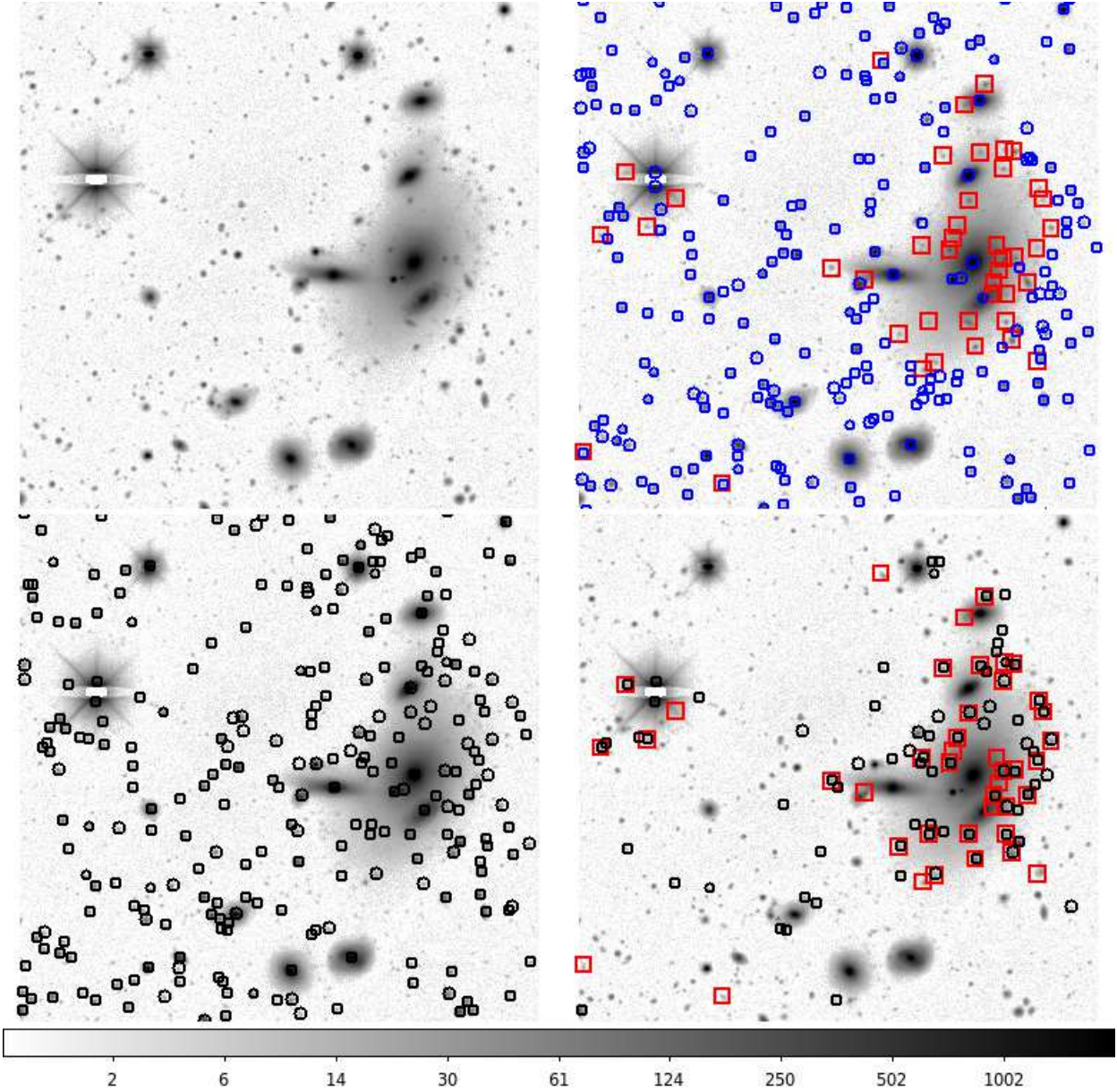}
\caption{ Upper Left: the linear combination of DES $r$, $i$ and $z$ coadd images for a cluster field. There is a bright star on the left and a brightest cluster galaxy on the right. Upper Right: same image with detected sources marked out. Blue circles are SExtractor detections (226 objects) using standard DES pipeline settings. Red squares (42 detections) are additional detections found through GAIN. Lower left: SExtractor detections with an aggressive deblending setting ($\mathrm{DEBLEND\_MINCONT}$ = $1\times 10^{-6}$), which tends to introduce many spurious detections. Lower Right: black circles (80 detections) are the additional Sextractor detections gained by more aggressive deblending. For comparison, GAIN detections from the upper right panel are shown again in this figure. In these figures, we only show sources brighter than 24.0 $\mathrm{mag}$ in $i$.}
\label{fig:app1}
\end{figure*}

\begin{figure*}
\includegraphics[width=1.0\textwidth]{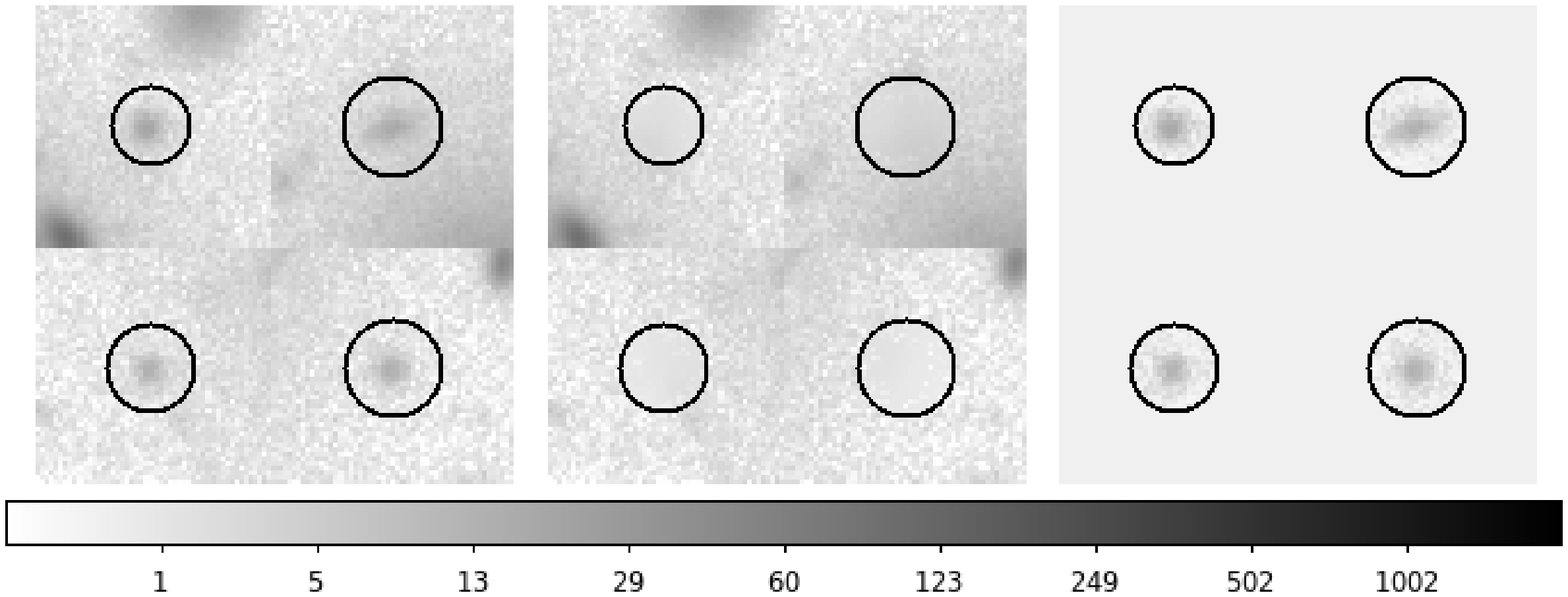}
\caption{ Left: r band DES coadd images of four blended sources. The circles indicate the areas to be deblended. Middle: images that contain light from neighbors of the blended sources. Right: the residual between the left and middle panels, which shows the light from the blended sources within the circle. }
\label{fig:app2}
\end{figure*}

The Gradient And INterpolation based deblender (GAIN deblender) is written in c++ and IDL and can be acquired online\footnote{\texttt{https://github.com/yyzhang/gain\_deblend  }}. It operates on processed images that have been cleaned: flatfielded, background subtracted, with cosmic rays removed. We assume that the users have already done one round of source extraction by other means, and are using this package for finding blended sources that are missed in the previous procedure. The final output from GAIN are deblended (blending-effect removed) images of blended sources. GAIN does not include any catalog extraction module for producing catalogs from images. In our application in this paper, we use SExtractor  to produce source catalogs from the deblended images.

The first function of the software is the identification of blended sources. It begins by independently identifying all sources in the image, then matching its results to the user supplied object list. Sources not matched to the user supplied input list are kept as new sources to be extracted from the blended image. This procedure is illustrated in Figure~\ref{fig:app1}. The upper left panel in Figure~\ref{fig:app1} shows a DES image containing a bright star and a brightest cluster galaxy. The upper right panel of Figure~\ref{fig:app1} shows sources detected by SExtractor (226 objects in blue circles) and some additional sources identified by GAIN (42 objects in red boxes). Note that this package is sensitive to image imperfections as is the case around the saturated star in the left half of the image. In general, the GAIN algorithm can find blended sources without introducing as many false detections as SExtractor with an aggressive  deblending setting. This is illustrated by the 80 objects in black circles in the lower right panel of Figure~\ref{fig:app1}.

The second function of the software aims to correctly assign the light in each pixel to the blended sources. This procedure is illustrated in Figure~\ref{fig:app2}. Given a detected blended source, a region to be deblended is computed by our package. GAIN then interpolates for the light that comes from ``background sources'' as shown in the middle panel. The residual between the original image and the ``background sources'' interpolation is the light from the blended source alone (right panel). This constitutes the final output from GAIN. Users will need to employ an independent catalog-extraction software to construct catalogs from the deblended images.

We find GAIN to be useful for two scenarios. In many applications, GAIN would be used both to detect blended sources using the first component and to extract light for those sources using the second component. It may also be used by skipping the first step, and using the second component to extract light for user supplied  blended sources.  This step is recommended if the user has blended sources in their first round of source extraction and wants to obtain consistent photometry for all blended sources. 

The GAIN package is fast enough to apply to wide field optical surveys. On a workstation computer equipped with Intel Xeon Processor E5645, running on a $10,000\times 10,000$ pixels image without parallelization, the source detection module takes approximately 300 seconds to identify 10,000 sources (without matching to a user supplied input catalog). The light separation module takes approximately 10 seconds for every 1,000 sources. The memory usage of this package depends on the size of the input image. It is generally more than twice the size of the image. For example, the total size of four tiles of DES coadd images is $\sim3.5~\mathrm{GB}$,  the peak memory usage of GAIN  running on these images  can be higher than $8\mathrm{GB}$.

\section{Software Algorithms}
\label{sec:method}

\begin{figure*}
\includegraphics[width=1.0\textwidth]{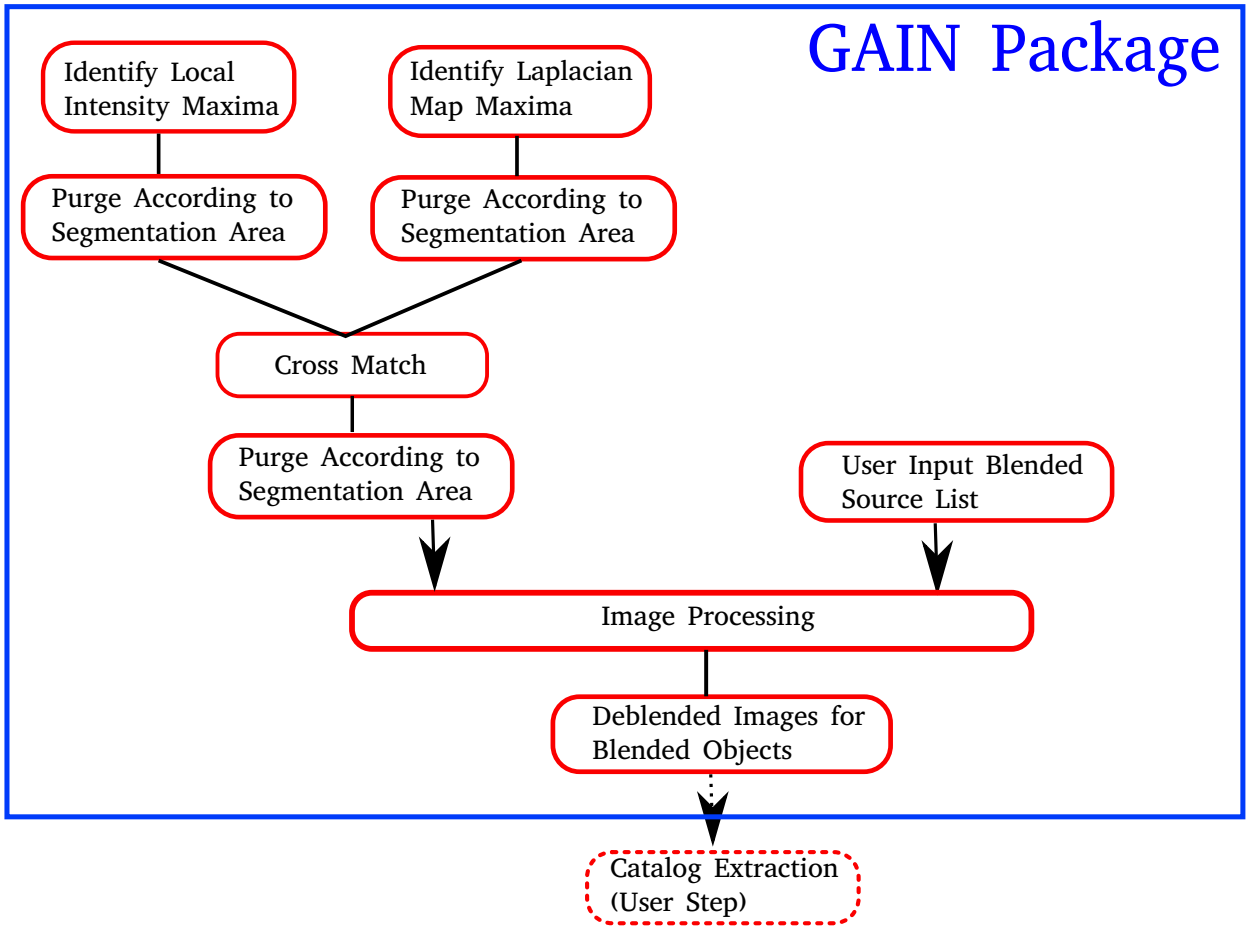}
\caption{A flow chart demonstrating the GAIN computing steps explained in Section~\ref{sec:method}. }
\label{fig:flowchart}
\end{figure*}

In the previous section, we mention that GAIN has two functions that can be combined for detecting blended sources and processing their images. In this section, we explain the algorithms used in these functions. The flowchart in Figure~\ref{fig:flowchart} demonstrates the computing steps during the application of the two functions. 

\subsection{Source Detection}
\label{sec:source}

The source detection component of GAIN aims to identify the local maxima in images which are associated with real objects. Our approach is inspired by the crowded field stellar photometry software DAOPHOT \citep{1987PASP...99..191S}. The presence of a separable astronomical object usually causes a local image intensity maximum. Unfortunately, many, even most, local intensity maxima are generated by image noise rather than real astronomical objects. Procedures to eliminate these noise peaks are therefore necessary.

One approach to reducing the impact of noise is to smooth the whole image. While smoothing is very effective at eliminating noise, it also washes out the saddle points which separate close pairs of real astronomical sources, exacerbating the problem of blending. When close pairs are separated by distances around twice the $\mathrm{FWHM}$ of the seeing, and one source is brighter than the other, even very modest smoothing merges the two. To do the best job of deblending, we would like to avoid smoothing. Instead of finding maxima in smoothed images, GAIN uses the image segmentation procedure and the image Laplacian map to reduce the impact of noise.

Source detection in GAIN begins with identifying sources on raw, unsmoothed images, then purges the identifications assigned with low pixel area in the segmentation map. To further eliminate spurious detections, GAIN cross matches the remaining identifications to sources identified in a ``weighted Laplacian'' map (which we explain in Section \ref{sec:lapmax}). The software performs two other rounds of segmentation area purging during the ``weighted Laplacian'' step and during the ``cross matching'' step.

\begin{figure*}
\includegraphics[width=1.0\textwidth]{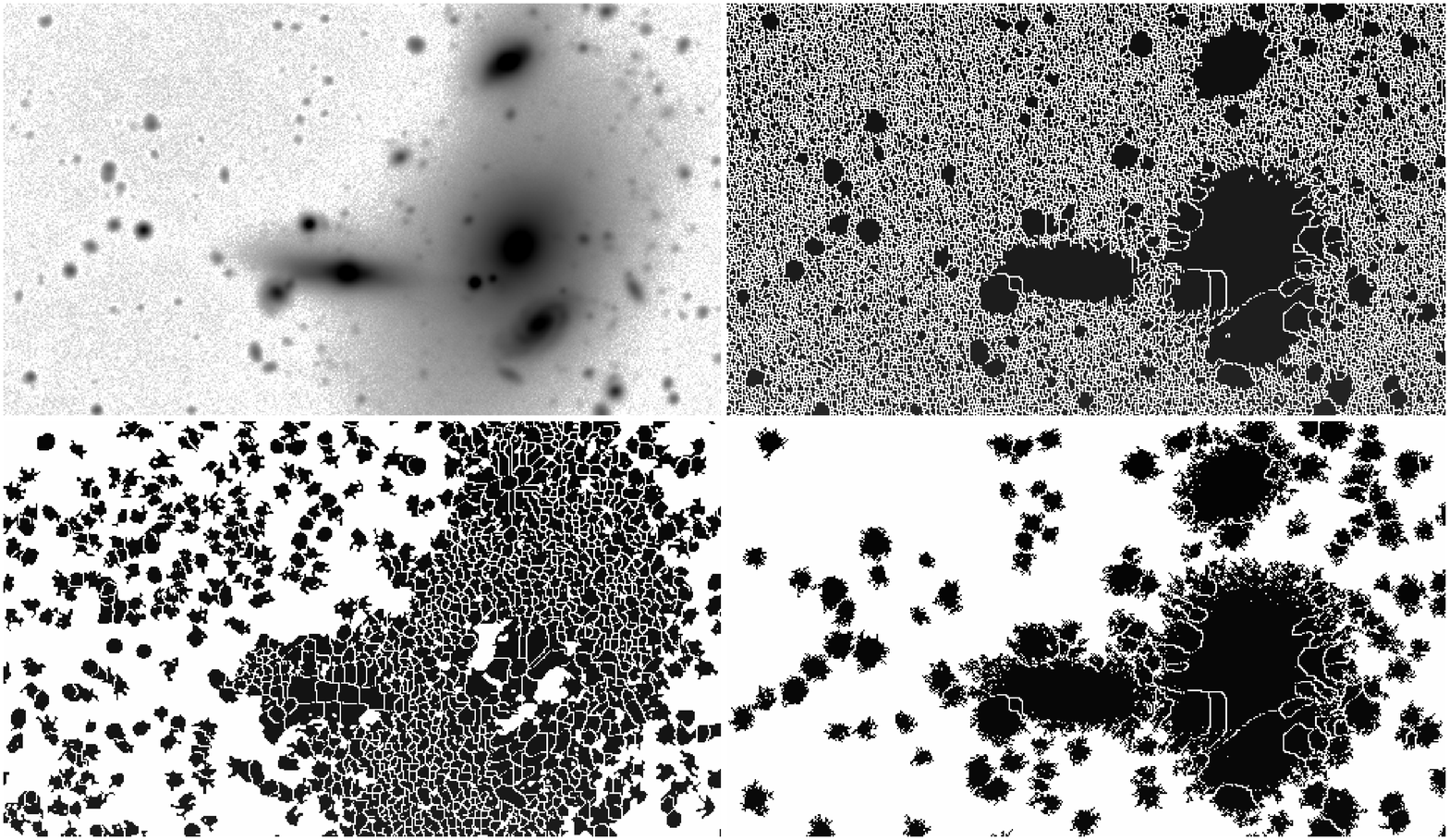}
\caption{ Upper left:  Original DES coadd image, same as the one shown in Figure~\ref{fig:app1}, but is zoomed in to best illustrate the segmentation procedure. Upper right: the segmentation map derived using all the local intensity maxima. Lower left: the segmentation map derived using all the local maxima of the weighted Laplacian map. Lower right: the segmentation map derived using the cross matched local maxima between the intensity map and the weighted Laplacian map. In the three segmentation maps, the empty white regions are the ``boundarys'' (see Section~\ref{sec:purge} for definition), and the pixels in black are associated with the seeds (local maxima from the intensity map, from the weighted laplacian map, and the cross matching between the two). }
\label{fig:seg}
\end{figure*}

\subsubsection{Segmentation of Images}
\label{sec:purge}

Purging of noise peaks can be aided by segmenting an image into separate regions, each associated with one element in the seed list of maxima. This seed list may include \textbf{all} local maxima, or may be produced by some other means, for example it could be a list of objects identified by SExtractor. In this first round of segmentation, we use image local intensity maxima, computationally defined as the pixels with intensity higher than the eight neighboring pixels. To segment the image, we use a simplified version of Meyer's watershed flooding algorithm \citep{785528}. To prime the segmentation procedure, we give each of the pixels in the seed list a unique region label. The goal of segmentation is then to label every pixel in the image as either belonging to one of the regions or residing in a boundary between the regions. The process begins by ranking all pixels in the image in descending order of intensity. As we move down the list, we apply the following procedure to each pixel.

\begin{enumerate}
\item If this pixel is already labeled (because it was in the seed list), its label remains unchanged.
\item If this pixel is unlabeled, and all of the neighboring pixels are unlabeled, this pixel is marked as a {\it boundary}. This should be true only for local maxima not found in the seed list.
\item If this pixel is unlabeled, and some or all of the neighboring pixels are already labeled, and their labels (except those labeled as {\it boundary}) are {\bf not} all the same, this pixel is labeled as a {\it boundary}.
\item If this pixel is unlabeled, and some or all of the neighboring pixels are already labeled, and their labels (except those labeled as {\it boundary}) are all the same, this pixel is given that label. If all the labeled neighbors are labeled as {\it boundary}, this pixel will also be a {\it boundary}. 
\end{enumerate}

Note that this procedure differs from the original implementation of Meyer's algorithm. When the original seed list includes {\bf all} of the local maxima (as shown in the upper right panel of Figure~\ref{fig:seg}), the procedure above gives the same result as Meyer's, though it is much faster. When the seed list does not include all of the local maxima, as will be the case in our GAIN application, this algorithm yields thick boundaries (shown in the lower panels of Figure~\ref{fig:seg}). This may not be desirable for computer vision applications like those for which Meyer's watershed algorithm was invented, but it is acceptable for our purposes.

After the segmentation is complete, we count the number of pixels labeled as belonging to each element of the original seed list. This number reflects the total area associated with each seed by the algorithm. We then purge false detections by eliminating seeds with segmentation areas smaller than an adjustable threshold value. For example, we purge all seed maxima associated with fewer than 27 pixels in DES coadd images. This threshold is set to reflect the the typical size of unresolved point sources in DES coadd images, but its value is finalized through trial and error, as are the other threshold values mentioned in this paper.

\subsubsection{Cross Matching to Laplacian Maxima}
\label{sec:lapmax}

\begin{figure*}
\includegraphics[width=1.0\textwidth]{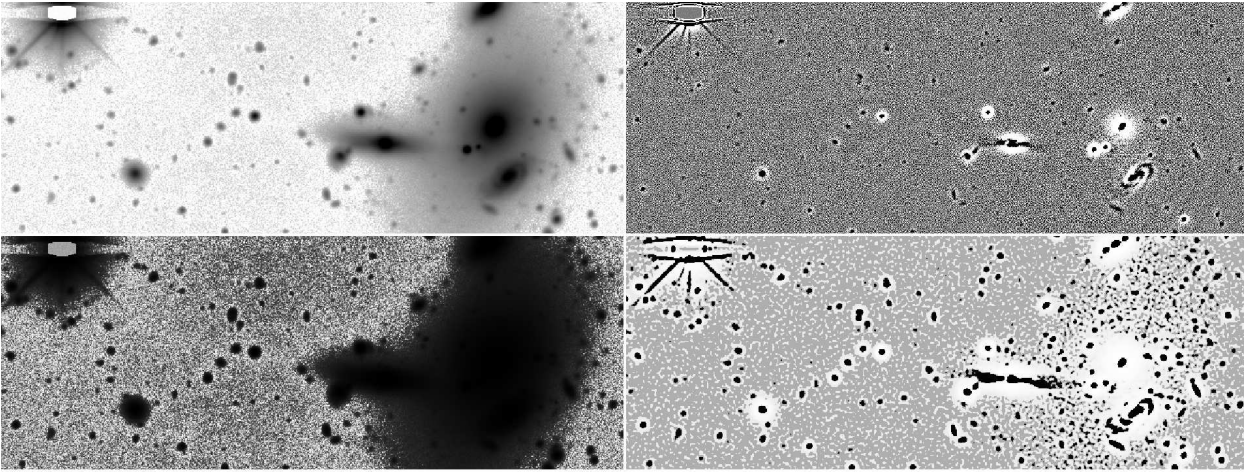}
\caption{ Upper Left:  original DES coadd image, same as the one shown in Figure~\ref{fig:app1} and Figure~\ref{fig:seg}. Upper Right: the Laplacian map of the same image. The fine features of the original image, like the spiral arms of the galaxy at the bottom right, are enhanced in the Laplacian map. Lower Left: intensity weighting map ($I_w$) derived as described in Section~\ref{sec:lapmax}. High intensity peaks are suppressed in this weighting map. Lower right: weighted Laplacian map in Section ~\ref{sec:lapmax} which combines the Laplacian map with  the intensity weighting map to bring out faint features.}
\label{fig:lap}
\end{figure*}

To populate an effective list of sources for deblending, GAIN generates a Laplacian map ($3\times3$ pixels Laplacian) of the original image. This is a useful approach because the contrast between real objects and their local background can be greatly enhanced in a Laplacian map. In terrestrial image processing applications, the Laplacian of Gaussian method is often used for edge detection  \citep{1994sstc.book.....L}. In these applications, an image is first smoothed on some scale with a Gaussian kernel, and then the Laplacian of the resulting image is calculated. However, in astronomical images, real objects already have had their spatial extent smoothed by an instrumental PSF, so that finding features in the Laplacian is an effective approach to object detection.

One limitation of this approach is that extended low contrast sources are not prominent in the Laplacian map. Since many of the sources indeed have low surface-brightness, we ameliorate this problem by weighting the Laplacian map with a transformed version of the original image intensity map. This weight map, denoted as $\mathrm{I}_w(x, y)$, is computed from the original image intensity map $\mathrm{I}(x, y)$ as,

\begin{equation}
\begin{split}
\mathrm{I'}(x, y)&=\mathrm{I}(x, y)-\mathrm{min}(\mathrm{I}) \\
\mathrm{I}_w(x, y)&=
 \left\{ \begin{array}{rl}
 \mathrm{I'}(x, y) , \mbox{ if $\mathrm{I'}(x, y)\leq \mathrm{mean}(\mathrm{I'})$} \\
 \log \frac{\mathrm{I'}(x, y)}{\mathrm{mean}(\mathrm{I'})} +\mathrm{mean}(\mathrm{I'}), \mbox{ otherwise.}
 \end{array} \right.
\end{split}
\end{equation}

We use the logarithmic values for high intensity pixels to suppress their weight. The weighted Laplacian map is then derived from the raw Laplacian map $\mathrm{L}(x, y)$ and the image weight map $\mathrm{I}_w(x, y)$, as
\begin{equation}
\mathrm{L}_w(x, y)=\mathrm{L}(x, y)\mathrm{I}_w(x, y).
\end{equation}

Unlike the original image intensity map, where pixel values range across many orders of magnitude, the pixel values of $\mathrm{I}_w(x, y)$ span a much narrower range. Finally, we smooth $\mathrm{L}_w(x, y)$ with a Gaussian function ($\sigma=1\,\mathrm{pixel}$ for application to DES coadd images, value adjustable). Local maxima identified in this smoothed, weighted Laplacian image then become the seeds for the segmentation and purging step described in Section~\ref{sec:purge}. For illustration, we show the smoothed, weighted Laplacian image of a DES coadd image in Figure~\ref{fig:lap}. We also show the segmentation result using local maxima from this smoothed, weighted Laplacian image in Figure~\ref{fig:seg}.

\subsubsection{Further Purging}

As a final step of cleaning, we cross match the complete list of local maxima from the original image with the list of local maxima from the smoothed and weighted Laplacian image. We retain those local maxima which have corresponding peaks (within 2 pixels or  0.53'') in the smoothed and weighted Laplacian image. This list of matched local maxima then becomes the seed list for the segmentation process (illustrated in the lower right panel of Figure~\ref{fig:seg}). Local maxima associated with sufficiently large areas in the segmentation map then form the final list of GAIN detected objects.

\subsubsection{Matching to User Supplied Catalog}

The usual application of GAIN follows an initial round of source extraction using tools like SExtractor. GAIN aims to search for additional blended sources missed by these applications. For this reason, we match the GAIN source list to the user supplied source list, and single out those sources not identified by the original reduction as a list of newly identified, deblended sources. 

For this matching procedure, the $(\mathrm{x}, \mathrm{y})$ coordinates of the user supplied sources are taken as input. Then in descending order of the intensity value at the user supplied sources' coordinates, we search for each source's nearest match in our source list from Section~\ref{sec:source}. All matches with a separation less than a threshold value (set to 10 pixels or 2.7'' for our application to the DES data) are considered valid matches. The matched source is then removed from our list and the matching process continues. After removing all GAIN sources which match the user supplied list, the remaining sources constitute our list of newly identified, deblended sources.

\subsection{Deblend Blended Sources}

The most difficult aspect of blended object photometry is untangling the relative contributions of light from multiple sources. It is not uncommon for a faint source near a bright object to sit atop a background with photon count equal to -- or in extreme cases several times higher than -- the photon count of the source. To measure photometry for blended objects, we must account for the light contributed by their neighbors.

\subsubsection{Separating the Light}

To disentangle light from multiple sources, we use an image inpainting technique originally developed in the computer vision field.  In this field, many techniques have been developed to recover damaged parts of an image, or to remove components that are unwanted. Our method is inspired by the \citet{doi:10.1080/10867651.2004.10487596} technique which is used to ``inpaint'' an image, i.e., to recover the texture of a small patch of an image from its surrounding pixels. It cannot re-create new patterns in the to-be-filled region, but rather fills them with a smooth background through interpolation. The problem this technique tries to solve is similar to our light separation problem. It allows us to estimate the light contribution from the more extended sources in the blended pixels. We choose the \citet{doi:10.1080/10867651.2004.10487596} technique over a variety of other available approaches \citep{Bertalmio:2000:II:344779.344972, 1323101} because it is computationally efficient and has been extensively studied.

The \citet{doi:10.1080/10867651.2004.10487596} technique works as follows. Given an intensity map, an unknown pixel can be inpainted with a value approximated from its known neighbors. \citet{doi:10.1080/10867651.2004.10487596} developed an efficient way to prioritize pixels in an unknown patch and determine the order in which they are inpainted, starting from the pixels nearest to the boundary and progressing inward \citep{Sethian20021996}. This technique explicitly maintains a {\it narrow band} of pixels to be filled in as one of its features. Our implementation is adapted from the \citet{doi:10.1080/10867651.2004.10487596} technique, with  a few modifications. A brief description follows.
\begin{enumerate}
\item Identify the regions to be inpainted: for each deblended source, this region is defined as a circle centered on the object, with an area equal to the segmentation area from Section~\ref{sec:purge}. Pixels in this region are labeled as {\it unknown}, and the rest as {\it known}.
\item \label{item:itemnb} Initiate the {\it narrow band}: the {\it narrow band} is a list of pixels originally identified as {\it unknown}, that have at least one neighbor labeled as {\it known}. Pixels in the {\it narrow band} are prioritized in the ascending order of their original intensity value.
\item Begin inpainting: select the highest priority pixel from the {\it narrow band} and inpaint it. We explain how this is done in the next two items. After inpainting this pixel, label it as {\it known}, and check if it has any {\it unknown} neighbors. If there are any, add them to the {\it narrow band} list, re-prioritize the {\it narrow band}, and repeat this step until the {\it narrow band} is empty.
\item \label{item:inpaint} Inpaint a pixel: to inpaint one pixel, we fill it with the zeroth order approximation value from its known neighbors,
\begin{equation}
\mathrm{I}(q)=\frac{\Sigma_p w_p \times \mathrm{I}(p)}{\Sigma_p w_p } 
\label{eq:zeroth}
\end{equation}
with $w_p$ being the weighting for each neighboring pixel.
\item The computation of $w_p$ follows the original definition in \citet{doi:10.1080/10867651.2004.10487596} as 
\begin{equation}
w_p=dir(p, q)\cdot dst(p, q) \cdot lev(p, q) 
\label{eq:wp}
\end{equation} with
\begin{equation}
\begin{split}
dir(p, q)& = \frac{ (\vec{p}-\vec{q}) }{ ||\vec{p}-\vec{q}|| } \cdot \vec{N}(p), \,  \vec{N}(p) = \nabla{I(p)}, \\
dst(p, q) & = \frac{1}{||\vec{p}-\vec{q}|| ^2} ~ \mathrm{and}\\
lev(p, q) & = \frac{1}{1+|I_\mathrm{orginal}(p)-I_\mathrm{orginal}(q)|}.
\end{split}
\label{eq:wp_comp}
\end{equation}
Here, $ \vec{p}-\vec{q} $ is the vector from pixel $p$ to pixel $q$. $dir(p, q)$ ($dir$ stands for {\it direction}) evaluates if the $p$ and $q$ pixels are aligned with the image intensity gradient direction $\vec{N}(p)$. The gradient vector, $\vec{N}(p) = \nabla{I(p)}$, is approximated with the four neighboring pixels of $p$. $dst(p, q)$ ($dst$ stands for {\it distance}) evaluates the distance between $p$ and $q$. $lev(p, q)$ ($lev$ stands for {\it level}) evaluates the closeness of image intensity before the inpainting procedure. 
\end{enumerate}

At step~\ref{item:itemnb}, we prioritize the {\it narrow band} pixels according to their image intensity values. This deviates from the fundamental feature of  the \citet{doi:10.1080/10867651.2004.10487596} algorithm in that \citet{doi:10.1080/10867651.2004.10487596} prioritize the {\it narrow band} in the order of pixels' distance to the original {\it known} and {\it unknown} region boundary, while the distance to the boundary is calculated from the fast marching solution to the Eikonal Equation \citep{Sethian20021996}. This is designed to mimic the practice in manual inpainting that the pixels closest to the {\it known} region are filled first \citep{Bertalmio:2000:II:344779.344972, 990497}. However, we prioritize pixels for inpainting using their intensity value rather than their distance to the {\it known} region. This is because pixels with lower intensity are less affected by the astronomical object we are removing, and filling them in first allows for more reliable background reconstruction. Also, we use Equation~\ref{eq:zeroth} for step~\ref{item:inpaint} with zeroth order approximation rather than the first order approximation used in \cite{doi:10.1080/10867651.2004.10487596} because astronomical images are noisy, and the derivatives at pixel scale are unreliable.

After using the above method for ``background'' interpolation for one object, the interpolated image contains light from the object's neighbors. The difference between the original image and the interpolated image contains the extracted light for this deblended object.

\subsubsection{Catalog Production}
\label{sec:cat}

GAIN does not contain a module that produces source catalogs from deblended images.
To produce a useful catalog of deblended sources, we need to measure magnitudes and shapes, as well as to classify each as a star or a galaxy. Many packages capable of doing this are available \citep{1996A&AS..117..393B, 0067-0049-142-1-1, 2002AJ....124..266P, 2010AJ....139.2097P}, and a user might choose their favorite. For our application to the DES data, we use SExtractor \citep{1996A&AS..117..393B}. The tests described in the following sections are also based on the application of SExtractor. We provide the wrapper code for such an application in our package. Applications of other software, like GALFIT\citep{2002AJ....124..266P, 2010AJ....139.2097P}, are possible as well. 

When choosing software for cataloging, we advise users to consider a few things:

\begin{enumerate}
\item The light extracted image of one object may be smaller than the area that contains {\bf all} of its light. One should consider how to reconstruct/account for the light of the object outside this region.
\item Because of the above constraint, photometry from model fitting is probably more appropriate. With our application of SExtractor, we find that Kron \citep{1980ApJS...43..305K} and Petrosian \citep{1976ApJ...209L...1P} magnitudes can provide reliable photometry.
\item Some star/galaxy separation methods may not work on light extracted images. When we use SExtractor on such images, we find that the class\_star quantity fails most of the time because of the small area used for light separation. Star/galaxy separation using SExtractor spread\_model quantity \citep[a classifier that evaluates object profile with local PSF,][]{2011ASPC..442..435B} is still effective.
\end{enumerate}

\section{Methods Validation}
\label{sec:veri}
\begin{figure*}
\includegraphics[width=1.0\textwidth]{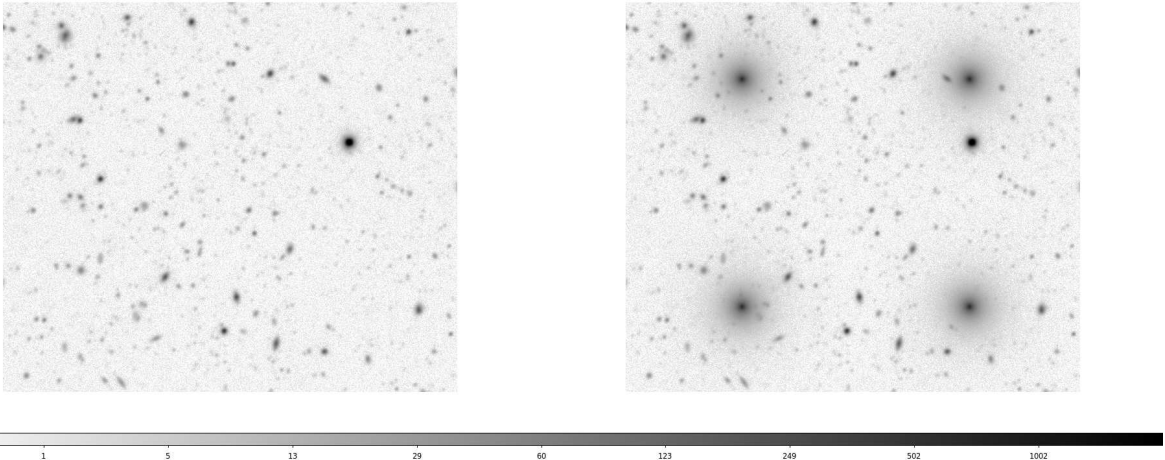}
\caption{ Left: An r band DES coadd image. The 10 sigma limiting magnitude of this image is $25.3\,\mathrm{mag}$ as measured from the SExtractor $\mathrm{mag\_auto}$ uncertainty. Right: The same image after adding simulated BCGs. The size of these two images is approximately $3.15\mathrm{'}\times3.15\mathrm{'}$. The apparent magnitude of the simulated BCGs is approximately $\sim$ 19 $\mathrm{mag}$.}
\label{fig:illu}
\end{figure*}

We verify the performance of the following aspects of GAIN : photometry measurement, source detection completeness, and source detection purity. We also include a modest test on star/galaxy separation as part of the photometry test.

We want to test this package on optical images with complex deblending challenges, while all the sources in these images are known and already reliably measured. Our principal goal is to improve deblending around bright cluster galaxies, so we designed a test to simulate this challenge  by adding simulated Brightest Cluster Galaxies (BCGs) to real deep optical images. For this test, we use deep coadd images from DES. Cleaning of individual exposures, coaddition of the images, and initial extraction of sources are all done using standard data processing pipelines from the DES collaboration \citep{2012SPIE.8451E..0DM}. We then select regions with few bright stars or real BCGs, so that deblending is not an important issue before the addition of a simulated BCG. Object catalogs extracted from these images are then used as ``truth tables'' in our testing procedure. When we add simulated BCGs to these images, some objects which are initially isolated and clean become blended, giving us a well understood deblending challenge to test against.

In Figure~\ref{fig:illu}, we show an image before and after adding simulated BCGs. For the results presented in this section, we make simulated galaxy images of Sersic profile with Sersic index $\mathrm{n}=4$ and Sersic radius $\mathrm{Re}=10''$ at $\sim19.0$ magnitude (exact values vary depending on how brightness is measured). We convolve these profiles with the PSF function and add them into the image. We also ran the test with simulated galaxies of different magnitudes ($\pm 2$ magnitude) and different Sersic parameters. The results are qualitatively independent of these changes. In this test, we combine GAIN with SExtractor for photometry measurement. We compare the performance of this set-up to the result from solely using SExtractor. Throughout the test, the major SExtractor deblending parameter, $\mathrm{DEBLEND\_MINCONT}$ is set at $0.001$, which is found to be optimum for  processing Dark Energy Survey early data.

We note that the algorithm we describe in this paper is designed for deblending between a satellite object and its much brighter neighbors. It may also be desirable to deblend closely spaced pairs and triples of astronomical objects. GAIN can indeed help with this kind of deblending problem, but its performance in these applications remains unverified. Because pairs or triplets do not always cause local maxima, it is hard to distinguish them from extended sources without using the image PSF (which is being implemented in a future version of SExtractor). GAIN is not optimized for this kind of deblending.

Finally, to thoroughly evaluate the deblending problem, the performance of GAIN, and more importantly the detection and photometry measurement of cluster galaxies in DES, one may wish to compare a data set with much higher resolution to DES data. A project comparing HST and DES data to study the DES data processing performance and its scientific effects is in progress (Palmese et al., in prep.).

\subsection{Photometry Measurement}
\label{sec:photo}

\begin{figure*}
\includegraphics[width=1.0\textwidth]{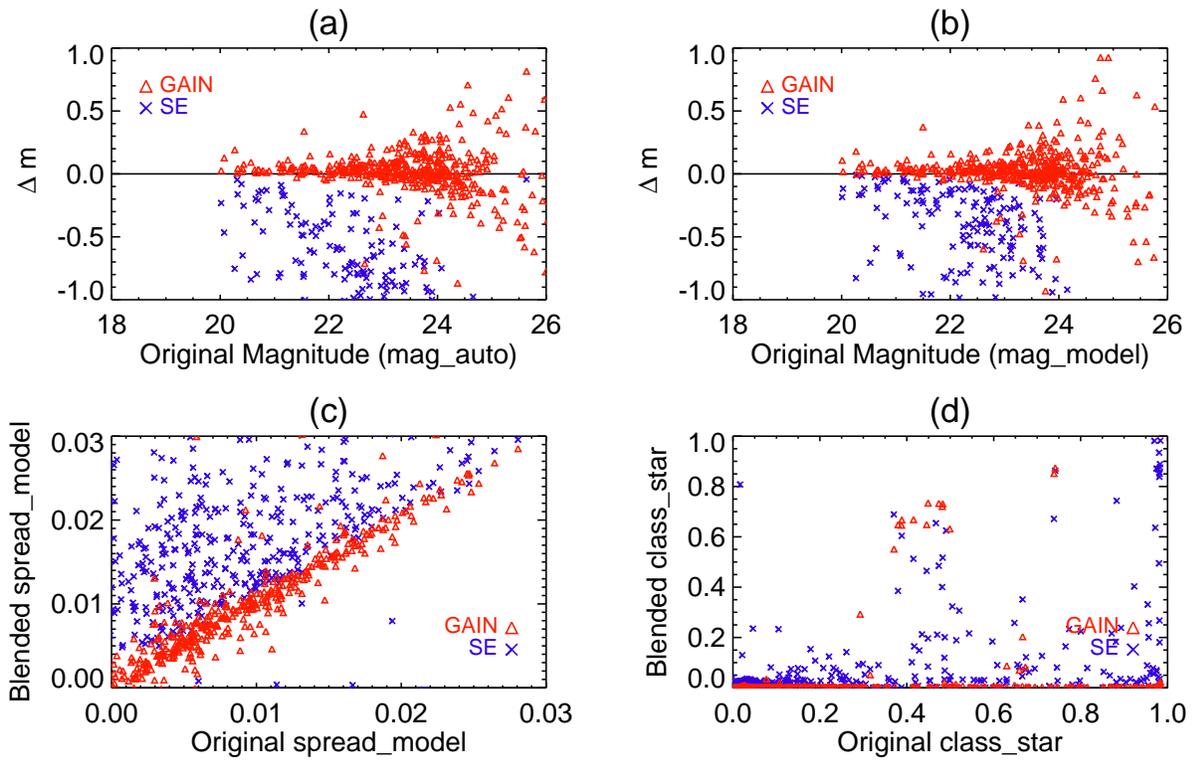}
\caption{ Comparison of photometry measurements and star/galaxy separation quantities with SExtractor  using the ``global background'' setting. (a)(b) Offsets between the "truth" magnitudes and measurements from altered images for ``artificially'' blended objects. (c)(d) Comparison of star/galaxy separation quantities for ``artificially'' blended objects.}
\label{fig:photo}
\end{figure*}

\begin{figure*}
\includegraphics[width=1.0\textwidth]{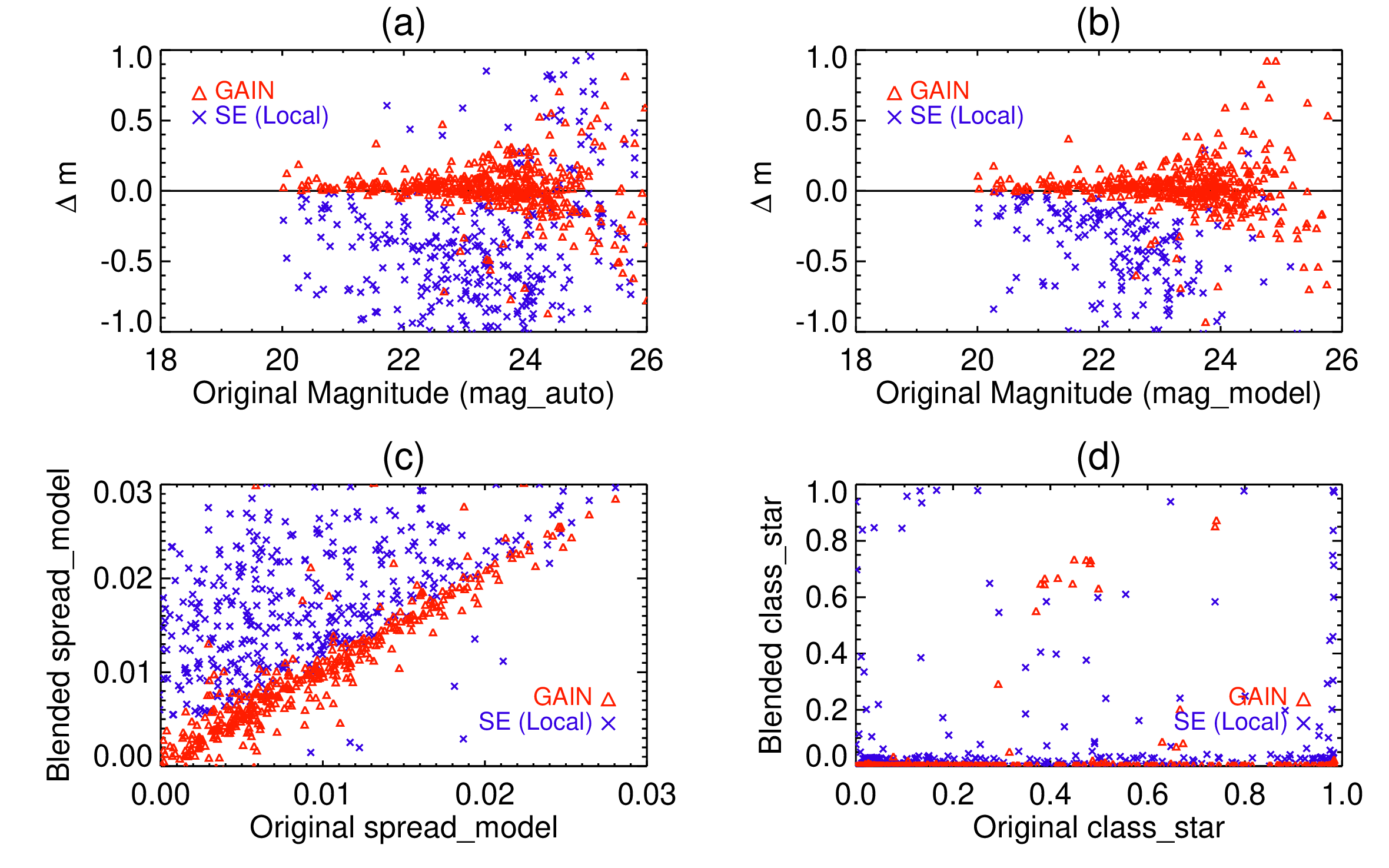}
\caption{Comparison of photometry measurements and star/galaxy separation quantities with SExtractor  using the ``local background'' setting. (a)(b) Offsets between "truth" magnitudes and measurements from altered images for ``artificially'' blended objects. (c)(d) Comparison of star/galaxy separation quantities for ``artificially'' blended objects.}
\label{fig:photolocal}
\end{figure*}

To test whether GAIN can improve photometry measurement for blended sources, we use it to measure sources that become blended in the altered coadd image. We run SExtractor on the unaltered image as well as the image with simulated BCGs. We select sources that are flagged to be unblended (SExtractor flag = 0) and isolated in the original image but become blended (SExtractor flag = 3) upon the addition of simulated BCGs. As these objects are considered to be clean in the original image, we treat their photometry measurement from this image to be the truth. In the images altered with simulated BCGs, we compare measurements of these sources from SExtractor directly and from SExtractor with GAIN implementation to their ``truth'' values. The result is shown in Figure~\ref{fig:photo} (a)(b).

Because light from simulated BCGs in the altered image is not completely accounted for in the basic SExtractor reductions, the blended sources typically have their brightness overestimated, often by as much as $0.5\,\mathrm{mag}$. Comparing to model magnitudes \citep{2011ASPC..442..435B, 2012ApJ...757...83D} in the truth table, Kron magnitudes ($\mathrm{mag\_auto}$) from the blended image are subject to more bias than model magnitudes. When GAIN is implemented, the measurements are significantly improved:  both model magnitudes and Kron magnitudes for these ``artificially'' blended sources appear to be unbiased. 

The photometry measurements from SExtractor in Figure~\ref{fig:photo} are obtained with the ``global background'' evaluation setting. In Figure~\ref{fig:photolocal}, we show comparisons adopting SExtractor local background setting. A local background setting does help diminishing the biases, but is not sufficient to eliminate them. In addition, the scatter of photometry measurement is much larger with a local background setting. 

In addition to magnitude measurements, we also include a modest comparison of star/galaxy separation parameters for these sources. This is shown in Figure~\ref{fig:photo} (c)(d) and Figure~\ref{fig:photolocal} (c)(d). We find that the $\mathrm{class\_star}$ quantity has become ineffective for star/galaxy separation as discussed in Section~\ref{sec:cat}. Another star/galaxy separation parameter $\mathrm{spread\_model}$ appears to remain effective.

\subsection{Purity and Completeness}

\begin{figure}
\includegraphics[width=0.5\textwidth]{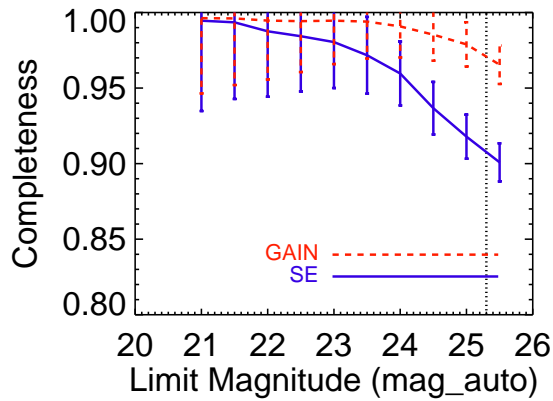}
\caption{ Completeness of  the C1 catalog (blue solid line, from SExtractor) and  completeness of the combination (red dashed line) of C1 and C2 (C2 from GAIN) as computed in Section~\ref{sec:comp}. The non-negligible incompleteness of SExtractor catalog can be improved by GAIN all the way to 25.5 $\mathrm{mag}$. The vertical dotted line shows the 10 sigma limiting magnitude of the image that the test is performed on. The errors in this plot are estimated assuming poisson distribution.}
\label{fig:com}
\end{figure}

\begin{figure}
\includegraphics[width=0.5\textwidth]{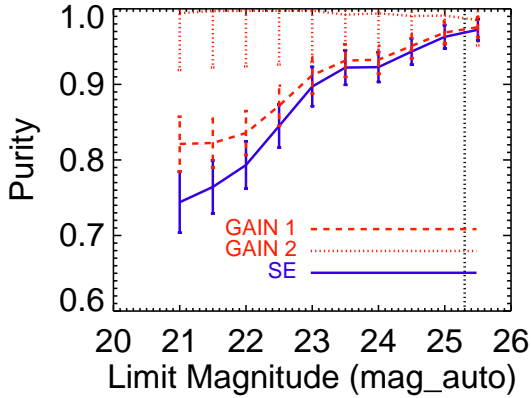}
\caption{ Purity of C1 (blue Solid line) by SExtractor and C2 (red dotted line) by GAIN and purity of the combination of C1 and C2 (red dashed line) as computed in Section~\ref{sec:pure}. The unsatisfying purity of C1 at the bright end indicates that the deblending procedure of SExtractor is prone to introduce spurious detections.  Also, the purity of the combination of C1 and C2 is affected by the purity of C1. On the other hand, the sources contained in C2 (GAIN output) are highly consistent with the previous running. The vertical dotted line marks the 10 sigma limiting magnitude of the image that the test is performed on. The errors in this plot are estimated assuming poisson distribution.}
\label{fig:pure}
\end{figure}

In this section, we examine the improvement of completeness (did we recover all real objects) and purity (are all the new deblended objects real) after implementing GAIN.

After introducing simulated BCGs into deep optical images, issues associated with the deblending procedures emerge: object detection becomes incomplete and spurious detections appear. We use GAIN to improve deblending and then examine its impact on detection completeness and purity. Note that the face values of completeness and purity presented in this section should not be taken as estimations for real astronomical images, as we are imposing exaggerated deblending difficulty in the testing images. The test here are only meant to show the effectiveness of GAIN with extreme situations.

\subsubsection{Completeness}
\label{sec:comp}

For the completeness test, we first run SExtractor on the unaltered coadd image, and use the resulting  catalog as the Truth Table 1 (TT1). Running GAIN on the original, unaltered coadd image produces a catalog of blended sources not detected in Truth Table 1, and we use this list of additional sources as Truth Table 2 (TT2). The combination of TT1 and TT2 is then used as the total truth table (TT) in our completeness and purity tests. We then insert simulated BCGs, run SExtractor on the image to produce Catalog 1 (C1), and run the GAIN package to search for blended sources and extract a supplemental Catalog 2 (C2). We then take all objects brighter than some magnitude limit from TT and match them to the C1 and C2 catalogs.

The matching is done in descending order of brightness for objects in the Truth Tables. For one object in the TTs, we search for the object that is nearest in C1 or C2. If the nearest neighbor from C1 or C2 is separated less than 5 pixels to the TT object,  we claim this object as {\it matched}. Once a C1/C2 object is used as a match, it is removed from the list available for matching. The TT sample is matched to the C1 or C2 sample deeper by $1 ~\mathrm{mag}$  to ensure that the completeness evaluation is not subject to photometry measurement scatters. We evaluate completeness by computing the ratio between the number of {\it matched} objects and the {\it total} number of objects in the TT sample. We compute this quantity for SExtractor by matching TT1 to C1 and for GAIN improved catalogs by matching TT to the combination of C1 and C2. The result is shown in Figure~\ref{fig:com}.

After the image is altered by simulated BCGs, a small but non-negligible fraction of sources are missed from SExtractor data reduction, especially at the faint end. The situation is noticeably improved after the application of GAIN, demonstrating the effectiveness of the software.

\subsubsection{Purity upon Deblending}
\label{sec:pure}

As there is no clear definition of ``real'' objects in the DES images,  for the purity test, we focus on evaluating the number of spurious detections introduced by the deblending procedure rather than categorizing objects as ``real'' or not.

To test the purity of the catalogs, we match C1 or C2 from the altered coadd images to Truth Tables from un-altered coadd images. The procedure is similar to the completeness test in Section~\ref{sec:comp}, except that we match Catalogs to the Truth Tables instead of  match Truth Tables to Catalogs.  We match the C1 or C2 sample above a magnitude limit to the TT sample deeper by 1$\mathrm{mag}$. We match C1 to T1, C2 to the combination of T1 and T2 and also the combination of C1 and C2 to the combination of T1 and T2. We calculate purity as the ratio between the number of {\it matched} objects in C1 or C2 sample and the {\it total} number of objects in the sample. The result is shown in Figure~\ref{fig:pure}. In this plot, GAIN 1 is the purity for the combination of C1 and C2, while GAIN2 is the purity of C2 alone. The purity of SExtractor (SE) is evaluated as the purity of C1 alone.

In Figure~\ref{fig:pure}, purity of C1 (SE) lowers toward the bright end, and is less than $80\%$ at magnitude 21. This is partially caused by deteriorated photometry and astrometry measurements of blended objects. Spurious and real objects in C1 are biased brighter because of blending (see discussion in ~\ref{sec:photo}), which affects the bright end of the purity test.  The purity of the combination of C1 and C2 (GAIN 1) is also negatively influenced by spurious detections in C1. For C2 (GAIN 2) alone, the sources contained in C2 are consistent with sources contained in T1 and T2 to $\sim99\%$. Figure~\ref{fig:pure} indicates that while performing the deblending procedure, SExtractor is likely to introduce spurious detections but GAIN is not.

\section{Discussion}

In this paper,  we describe modest tests showing that GAIN improves the reliability of photometry for blended objects. However, these tests are only designed to demonstrate the effectiveness of GAIN, and we recommend caution before applying these test results to new analyses. The completeness, purity, and photometry test results presented here should be considered valid only for this application of SExtractor and GAIN. GAIN is a supplement to object finding and photometry packages like SExtractor, and its performance inevitably depends on details of its image processing partner. In this paper we compare the SExtractor and GAIN combined photometry output to SExtractor output, but do not investigate the original SExtractor photometry measurements.

A key element in any discussion of galaxy photometry is the evaluation of background light, which mostly comes from the sky. No photometry measurement algorithm will produce reasonable results if the sky background is inaccurately estimated. For this reason, any tests of photometric reduction algorithms must address background subtraction. GAIN does not perform photometry measurements or sky subtraction, so its connection to this issue is remote. Nevertheless, we have tested GAIN's susceptibility to imperfect sky background subtraction by performing an additional round of background subtraction for the test images used in Section~\ref{sec:veri}. We employ the SExtractor local background evaluation function for this procedure, and then apply GAIN to these images that have went through an unnecessary (the images have already been background subtracted) and wrong (using improper SExtractor background evaluation setting) round of background subtraction. The results of GAIN application in this scenario remain quantitatively similar to those reported above, confirming our expectation that background subtraction has little impact on GAIN efficacy.

While the background estimation in this test is improper, it is not outrageously wrong. However, if it had been outrageously wrong, no photometry algorithm would produce reasonable results. This test at least shows that GAIN is not sensitive to imperfect background subtraction. Note that when performing image interpolation for blended sources, GAIN does perform an additional round of local background subtraction, in which the ``background'' light is contributed by neighboring objects. Our test in Section~\ref{sec:veri} have already verified GAIN's effectiveness on this aspect.

When applying GAIN for a specific science analysis, one may wish to design additional tests that focus on aspects of GAIN relevant to the analysis. For this paper, aimed at deblending galaxies in crowded cluster cores, the real image plus simulated galaxy approach outlined in Section 4 seems to be the most relevant approach. We also considered tests using simulated images, but found them more difficult to interpret. Completeness and purity measures vary with environment, and objectively distinguishing different environments is not trivial. Finally evaluating the realism of the simulated images without a specific scientific goal in mind is challenging. It is for these reasons that we have chosen to test GAIN with this real image plus simulated galaxy approach. 

Finally, GAIN is not a substitute for packages that would yield precision photometry measurement for blended objects \nocite{2012MNRAS.422..449B, 2013ApJS..206...10G, 2014A&A...572A..87D} (GALAPAGOS Barden et al. 2012; Galametz et al. 2013; GASPHOT  D'Onofrio et al. 2014). These packages tend to perform two rounds of SExtractor source extraction with the second round tuned to pick up faint blended sources. As the second round of source extraction tends to yield many spurious objects and biased photometry,  precision photometry fitting software needs to be employed to purge and refine the final catalog. Compared to these packages, the biggest advantage of GAIN is speed. The multipass approaches used on crowded fields are generally too slow for full wide field survey data, limited by the speed of precision photometry fitting. For example, the GASPHOT package is about 100 times slower than GAIN. It took GAIN ~7 days to run for 300 deg$^2$ DES science verification data on a small computer cluster, but it would take GASPHOT 1.92 years  for the same field. GAIN is strikingly efficient in terms of computing demands.

\section{Summary}
\label{sec:sum}

Deep astronomical images face deblending challenges, especially in the crowded cores of galaxy clusters. Current deblending algorithms are not optimized to handle this problem. To take full advantage of the opportunity offered by new surveys like the DES, we need better methods for extracting accurate galaxy lists in cluster cores. In this paper, we describe a relatively simple approach to sorting out blended sources in these crowded regions. The design of this GAIN package includes two innovative features.

\begin{enumerate}
\item This package makes use of the Laplacian of an intensity image for blended source detection. In deblending procedures, one of the biggest challenges occurs when the intensity contrast between blended sources is too low to trigger detection. In this paper, we have shown that this problem can be alleviated by measuring the image intensity gradient. The image intensity gradient is often used in the computer vision field to bringing out fine details of an image. Future astronomical data production software can make use of this information to help deblending.
\item This package uses an interpolation technique to separate blended light from multiple sources. This is an improvement comparing to two popular approaches: simply assigning pixels to blended sources which is inaccurate but computationally efficient, and simultaneously fitting profiles of multiple sources which is accurate but computationally inefficient. Our method provides a nice balance between accuracy and efficiency.
\end{enumerate}

We have tested this package on DES coadd images. Our tests show that it can increase the reliability of photometry for blended objects. It can also increase the completeness of blended source detection, while introducing only a modest number of spurious detections. Upon application to DES data, GAIN has been used to improve cluster galaxy detection and modeling of cluster central galaxy light profile. It is also possible to apply GAIN to HST and SDSS images.

\acknowledgements

The authors are pleased to acknowledge supports of U.S. Department of Energy grants no. DE-FG02-95ER40899 and DE-AC02-76SF00515, and the University of Michigan Rackham predoctoral fellowship. We thank members of the Dark Energy Survey Data Management project, members of the Dark Energy Survey Cluster Working Group,  members of the Dark Energy Survey Galaxy Evolution Working Group, and the University of Michigan Dark Energy Survey team for suggestions and encouragements that greatly facilitated the progress on this work. We thank Gary Bernstein, Ofer Lahav, Gus Evrard, Nacho Sevilla, Tom Diehl, and Alistair Walker for comments on the paper draft.  We use DES Science Verification in this paper. We are grateful for the extraordinary contributions of our CTIO colleagues and the DECam Construction, Commissioning and Science Verification teams in achieving the excellent instrument and telescope conditions that have made this work possible. The success of this project also relies critically on the expertise and dedication of the DES Data Management group.

Funding for the DES Projects has been provided by the U.S. Department of Energy, the U.S. National Science Foundation, the Ministry of Science and Education of Spain, 
the Science and Technology Facilities Council of the United Kingdom, the Higher Education Funding Council for England, the National Center for Supercomputing 
Applications at the University of Illinois at Urbana-Champaign, the Kavli Institute of Cosmological Physics at the University of Chicago, Financiadora de Estudos e Projetos, 
Funda{\c c}{\~a}o Carlos Chagas Filho de Amparo {\`a} Pesquisa do Estado do Rio de Janeiro, Conselho Nacional de Desenvolvimento Cient{\'i}fico e Tecnol{\'o}gico and 
the Minist{\'e}rio da Ci{\^e}ncia e Tecnologia, the Deutsche Forschungsgemeinschaft and the Collaborating Institutions in the Dark Energy Survey.

The Collaborating Institutions are Argonne National Laboratory, the University of California at Santa Cruz, the University of Cambridge, Centro de Investigaciones Energeticas, 
Medioambientales y Tecnologicas-Madrid, the University of Chicago, University College London, the DES-Brazil Consortium, the Eidgen{\"o}ssische Technische Hochschule (ETH) Z{\"u}rich, 
Fermi National Accelerator Laboratory, the University of Edinburgh, the University of Illinois at Urbana-Champaign, the Institut de Ciencies de l'Espai (IEEC/CSIC), 
the Institut de Fisica d'Altes Energies, Lawrence Berkeley National Laboratory, the Ludwig-Maximilians Universit{\"a}t and the associated Excellence Cluster Universe, 
the University of Michigan, the National Optical Astronomy Observatory, the University of Nottingham, The Ohio State University, the University of Pennsylvania, the University of Portsmouth, 
SLAC National Accelerator Laboratory, Stanford University, the University of Sussex, and Texas A\&M University.

This paper has gone through internal review by the DES collaboration. 

\bibliography{reference}

\end{document}